\documentclass[twocolumn,preprintnumbers,amsmath,amssymb]{revtex4}

\usepackage{graphicx}
\usepackage{dcolumn}
\usepackage{bm}
\usepackage{amssymb}
\usepackage{amsmath}
\usepackage{enumerate}
\usepackage{wrapfig,epsfig}
\usepackage{graphics}
\usepackage{shadow}
\usepackage[T1]{fontenc}
\usepackage{color}
\usepackage{array}
\usepackage{rotating}
\newcommand{\Pe}{\text{Pe}}
\newcommand{\gammap}{\dot{\gamma}}
\newcommand{\dd}{\text{d}}
\newcommand{\cste}{\text{cste}}

\begin{document}

\title{Transverse transport of solutes between co-flowing
pressure-driven streams for microfluidic studies of diffusion/reaction processes}
\author{Jean-Baptiste Salmon}
\affiliation{
jean-baptiste.salmon-exterieur@eu.rhodia.com\\
LOF, UMR~5258 CNRS--Rhodia--Bordeaux 1,178 avenue 
du Docteur Schweitzer, F-33608  Pessac cedex, FRANCE}
\author{Armand Ajdari }
\affiliation{Th\'eorie et Microfluidique, UMR 7083 CNRS-ESPCI, 10 rue Vauquelin,75005 Paris, FRANCE}

\begin{abstract} 
We consider a situation commonly encountered in microfluidics: 
two streams of miscible liquids are brought at a junction to flow side by side within a microchannel, 
allowing solutes to diffuse from one stream to the other and possibly react.  
We focus on two model problems: (i) the transverse transport of a single solute from a stream 
into the adjacent one, (ii) the transport of the product of a diffusion-controlled
chemical reaction between solutes originating from the two streams.
Our description is made general through a non-dimensionalized formulation
that incorporates both the parabolic Poiseuille velocity profile along the channel 
and thermal diffusion in the transverse direction.
Numerical analysis over a wide range of the streamwise coordinate $x$ reveal different regimes. 
Close to the top and the bottom walls of the microchannel, the extent of the diffusive zone 
follows three distinct power law regimes as $x$ is increased, characterized respectively by 
the exponents $1/2$, $1/3$ and $1/2$. Simple analytical arguments are proposed to account for these results. 
\end{abstract}
\maketitle

\section{Introduction}
\begin{figure}[ht!]
\begin{center}
\scalebox{1}{\includegraphics{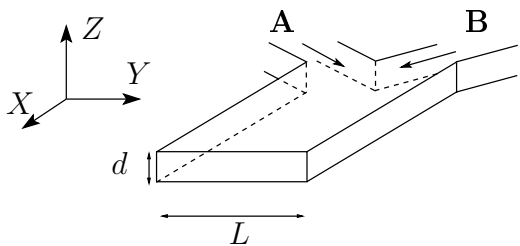}}
\caption{\small Geometry of the problem. The origin of the axes is set at the junction between the 
two inlets and
$Z$ ranges between $\pm d/2$. {\bf A} and~{\bf B} indicates the two arms  
in which the miscible liquids  are injected. The liquids  may react
when they interdiffuse into each other.} \label{setup}
\end{center}
\end{figure}

Microfluidics is a very promising format to measure the dynamics of different processes: diffusion of 
a solute \cite{Kamholz:01_1,Culbertson:02,Salmon:05}, protein folding \cite{Knight:98,Pollack:01,Pan:02},
kinetics of chemical reactions \cite{Pabit:02,Salmon:05_1}\dots\,
(see Refs.~\cite{Stone:04,Squires:05,Vilkner:04}  for reviews on microfluidics). 
In some of these approaches, one follows the reaction-diffusion dynamics of solutes between parallel streams.
Figure~\ref{setup} displays the simplest case of a Y-junction commonly encountered in microfluidic experiments, 
where two liquids are injected in a main microchannel.
Since flows are laminar given the small length scales and velocities involved, the 
liquids only mix by transverse diffusion, and one can relate the distance downstream the channel, 
to the time elapsed since the two streams where put into contact.  
Essentially, if the average velocity in the channel is $V$,
one expects that the situation at a distance $X$ from the junction
corresponds to the outcome of diffusion-reaction over a time $t=X/V$,
so that for example, simple interdiffusion leads to a diffusion zone of width
$Y \sim (2DX/V)^{1/2}$. 

 However, in confined geometries the situation is actually more complex as
pressure-driven flows are heterogenous (parabolic velocity profile), resulting
in residence times that vary with the distance to the walls.
Therefore, in order to extract from these experiments
physical parameters such as diffusion coefficients or rate constants, 
it is essential to understand correctly the mass transport phenomena in microchannels.

In this context, Ismagilov {\it et al.} \cite{Ismagilov:00} 
have studied using confocal microscopy, the formation and the diffusion of 
a fluorescent product in the diffusion {\it cone} of two reactive miscible solutions flowing 
side by side in a microchannel (the geometry is the same as displayed in Fig.~\ref{setup}). 
They have shown, both theoretically and experimentally, 
that the extent of the transverse diffusive zone near the top and the bottom walls ($Z = \pm d/2$) 
scales as $Y \sim X^{1/3}$, i.e. follows a $1/3$ power law in the streamwise coordinate~$X$. 
In the central plane of the microchannel, they observed 
the classical $1/2$ power law for diffusive processes $Y \sim X^{1/2}$.
 Such behaviours are due to the coupling of transverse diffusion and velocity gradients in the channel. These measured
exponents can be accounted for by arguments derived from the 
classical Lévêque problem~\cite{Stone:89}. 
Kamholz {\it et al.}  have performed numerical simulations of the same problem \cite{Kamholz:02}, 
and found the different regimes expected. However, their results
concern only specific sets of dimensionalized parameters which makes
it difficult to apply them to other experimental configurations.
In addition, some of the results displayed in Ref.~\cite{Kamholz:02} are inconsistent
with findings shown later in the present manuscript. 
More recently, a significant improvement has been brought by Jiménez who performed numerical simulations 
of the same problem using dimensionless equations\cite{Jimenez:05}. 
His work accounts for the different regimes observed experimentally, and the author also clarifies these results
using  
analytical solutions of the concentration profile in the range of small and large~$X$. 
In a recent paper  (Supporting Information of Ref.~\cite{Salmon:05_1}), 
we have studied the effect of the Poiseuille flow developped in a microchannel 
of high aspect ratio, on the reaction-diffusion dynamics 
occurring in the diffusion cone of two reactive solutions injected side by side. 
Through numerical simulations, we showed that for the high aspect ratio studied, 
a 2D description in terms of height-averaged concentrations and velocities
was appropriate and could be safely used to extract the rate constant of the reaction from experimental data.

Our aim in this paper is to revisit formally these transverse 
transport problems so as to provide generally applicable insights 
as to the outcome of experiments performed in arbitrary conditions
(flow rate, geometry,\dots), but for the requirement that the P\'eclet number be large (see below).
In the first part of this paper, we discuss the simple case of the transverse diffusion of a solute, 
using numerics and simple analytical arguments in a 
generic non-dimensionalized formulation. 
We clarify that the $1/3$ power law behaviour for transverse spreading 
($Y\sim X^{1/3}$) holds after the junction in layers close to the walls, 
the thickness of which increases as $\sim X^{1/3}$. For 
a small but finite distance from the walls (i.e. for a given $Z$), 
the transverse extent $Y$ of the diffusion zone actually follows successively 
three power law regimes in $X$ as $X$ is increased, with exponents $1/2$, then $1/3$ and finally $1/2$ again.
We then discuss the relevance of these findings for published data.
In a second part, we study the more complex case of coupled diffusion and reaction
 of two reactive solutes yielding the apparition and diffusion of a product
 in the neighbourhood of the interface between the two streams. Most of
 the phenomenology described in the first part is shown to persist
when the kinetics of the reaction is fast compared to diffusion,
with however a few qualitative differences that are pointed out.  

\section{Diffusion of a solute into a neighbour stream}
\subsection{Assumptions of the model}

We consider a microchannel 
with a high aspect ratio $\Gamma = L/d \gg 1$, as depicted in Fig.~\ref{setup}. 
Two solutions of the same solvent 
are injected at a same constant flow rate in the two arms of the microchannel: 
(i) the solution flown through arm {\bf A} contains at a dilute level a solute A, 
whereas (ii) pure solvent is flown through the other arm {\bf B}.
We are interested in the concentration profiles $A(X,Y,Z)$ of this solute downstream.
Given our microfluidic motivation, we consider that the flow is strictly laminar 
(small Reynolds number) and that 
the solute only disperses by molecular diffusion. 
We also suppose that the diffusion coefficient $D$ of the solute, as well as the 
fluid viscosity and density, are constant for the dilute solutions considered here. 

With these assumptions, the velocity profile is locally parabolic
and well described by Hele-Shaw formulas 
(except for positions $Y$ at a distance $d$ or smaller from the two lateral walls at $Y= \pm L/2$). 
Typically after an entrance region of order $L$ from the junction,
the flow takes the simple form ${\bf v}(X,Y,Z)= v(Z){\bf X}$ with
\begin{equation}
v(Z)= U\left(1 - \left(\frac{Z}{d/2}\right)^2\right)\,,
\end{equation}
with $U$ the maximal velocity.

We now write the equation describing the transport of the solute
resulting from convection in this velocity field and thermal diffusion.
We focus immediately on the limit of high P\'eclet numbers, i.e. $\Pe = U d /D \gg 1$, 
where one can neglect the diffusion along the flow direction, and
the concentration $A$ of solute A evolves according to
\begin{eqnarray}
v(Z)\partial_X A(X,Y,Z) = D(\partial_Y^2 + \partial_Z^2)A(X,Y,Z)\,. \label{eq:1}
\end{eqnarray}
If we neglect entrance effects, i.e. if we assume that the solute does not 
significantly diffuse transversely in the entrance zone,
the concentration downstream can be found by solving Eq.~(\ref{eq:1}) with 
no-flux boundary conditions on the lateral walls, and "initial" 
conditions at $X \simeq 0$: $A(Y,Z) = 0$ for $Y>0$ and $A(Y,Z) =A_0$ for $Y<0$. 

We make the problem non-dimensional using 
\begin{eqnarray}
x = X/(d\,\Pe), \,\, z = Z/d, \,\, y = Y/d,\,\, \text{and} \,\, a = A/A_0,\label{eq:adi} 
\end{eqnarray}
so that Eq.~(\ref{eq:1}) takes a parameter-free form
\begin{equation}
(1-(2z)^2)\,\partial_{x} a(x,y,z) = (\partial_y^2 + \partial_z^2)a(x,y,z)\,. \label{eq:principale} 
\end{equation}

\subsection{Numerical computation of the model}

We solve Eq.~(\ref{eq:principale}) using a simple numerical approach 
(the Euler method).
We descretize the $(y,z)$ plane using a 2D "grid" with nodes 
$y_j = -\Gamma/2 + \dd y/2 + (j-1)\dd y$, and $z_j = -1/2 + \dd z/2 + (j-1)\dd z$, 
with $\dd y = \Gamma/n$  and $\dd z = 1/p$, with typically $p=60$ points across the channel 
height and $n=80$ points across the channel width. 
We impose the classical no-flux boundary conditions at the walls at $z=\pm 1/2$ 
and $y = \pm \Gamma/2$.
As in Ref.~\cite{Jimenez:05}, 
we perform different simulations and overlap their outcome to cover a wide 
range of values of  $x$. More precisely, to reach sufficient accuracy over this whole range, 
the resolution on the $y$-axis is varied through variations of $\Gamma$, 
each simulations being terminated before the solute diffuses up to the side walls as $y = \pm \Gamma/2$.

\subsection{Numerical results}

\begin{figure}[ht!]
\begin{center}
\scalebox{1}{\includegraphics{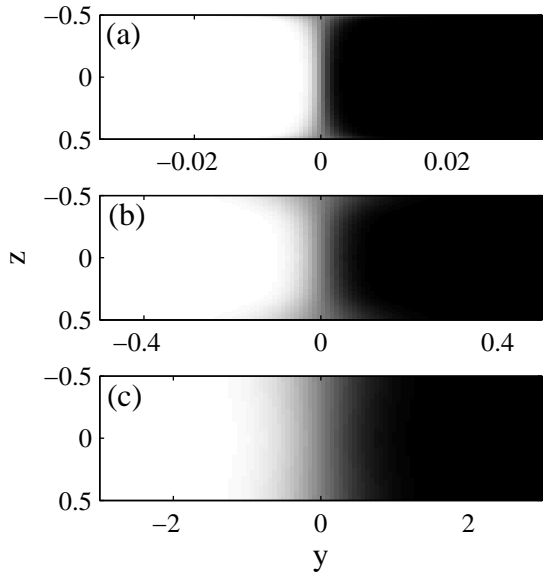}}
\end{center}
\caption{{\small Typical $z$-$y$ slices of the concentration $a$ of the solute at (a) 
$x=10^{-6}$, (b) $10^{-3}$, and (c) $10^{-1}$. A linear gray scale is used
to code the values of $a$: black corresponds to 0, and
white corresponds to 1.\label{cartes}}}
\end{figure}

Typical concentration profiles within cross-sections at different downstream locations $x$ are shown in Fig.~\ref{cartes}.
For $x\lesssim 0.1$, the concentration profiles 
are not homogenized across the height of the microchannel, as already explained 
in Refs.~\cite{Ismagilov:00,Kamholz:02,Jimenez:05}.
This is a consequence of the dispersion of the residence times in the microchannel 
due to the Poiseuille flow. For $x\gtrsim 0.1$, one almost recovers homogeneous profiles. 
This is expected since the criterion for the solute to sample all $z$-positions
by diffusion reads roughly $x\gtrsim 1/8$, as the solute has to diffuse over the half height of the channel 
so that $2 D X / V \gtrsim (d/2)^2$.

To get further insight into the broadening with $x$ of the concentration front, 
we define an average width $w(x,z)$ of the concentration profiles, 
at height $z$ and distance $x$, as
\begin{equation}
2w^2(x,z)= \int_{-\Gamma/2}^{\Gamma/2} \dd y\, y^2 \partial_y a(x,y,z) /\int_{-\Gamma/2}^{\Gamma/2} \dd y \, 
\partial_y a(x,y,z) \label{definition} \,. 
\end{equation}
In the ideal case of simple diffusion with a homogeneous velocity 
profile (plug flow), $a(x,y,z) = 0.5\left(1+\text{erf}(-y/(2w(x))\right)$, 
and the diffuse broadening obeys the simple diffusive scaling $w(x)\sim x^{1/2}$ with an exponent $1/2$.  

\begin{figure}[ht!]
\begin{center}
\scalebox{1}{\includegraphics{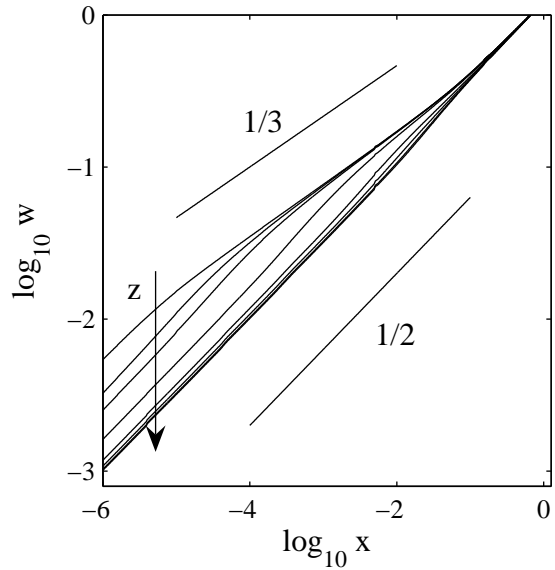}}
\end{center}
\caption{{\small  $\log_{10}(w)$ vs. $\log_{10}(x)$ for different $z$. The arrow indicates the range of increasing $z$
from $-1/2+ \dd z/2$ to $-\dd z /2$ ($\dd z = 1/60$ in the present simulations). 
The two continous lines indicate the $1/2$ and $1/3$ power law behaviours. \label{logwvslogx}}}
\end{figure}

The curves $w(x,z)$ vs. $x$ for a Poiseuille flow profile
are shown in Fig.~\ref{logwvslogx} for several heights $z$ in the microchannel.
For values of $x$ below 0.1, a dispersion of the widths $w(x,z)$ is visible, as expected from Fig.~\ref{cartes}: 
the concentration profiles are wider close to the bottom and the top walls. 
For values of $x$ greater than 0.1, all the curves collapse and a homogenous 
(along $z$) diffusion  along $y$ is recovered.

\begin{figure}[ht!]
\begin{center}
\scalebox{1}{\includegraphics{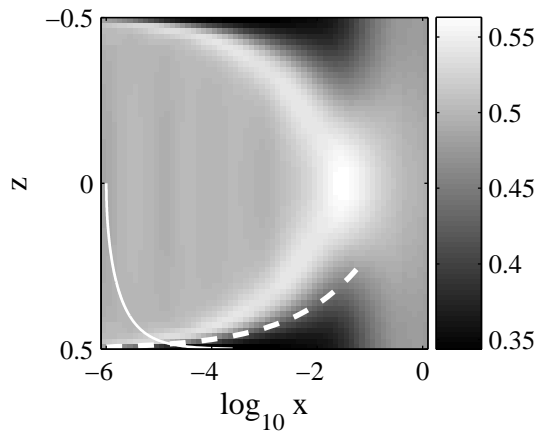}}
\end{center}
\caption{{\small  $\alpha(z,x)$ vs. $\log_{10}(x)$ and $z$. The exponents $\alpha(x,z)$
are estimated by local fits of the curves displayed in Fig.~\ref{logwvslogx} by power laws.
The dashed white line indicates $u=1$ [see Eqs.~(\ref{closetothewall})--(\ref{eq:armand})], and the continous white line 
corresponds to Eq.~(\ref{eq:validity}) with $\Pe = 1000$. For clarity, these two lines have been
only plotted for $z>0$. \label{alphazxvslogx}}}
\end{figure}

To describe in more detail the {\it local} behaviour of $w(x,z)$ vs. $x$, 
we fit these curves locally by power laws
$w(x,z) \sim x^{\alpha(x,z)}$ over a 
{\it sliding} decade of $x$, from $x\approx10^{-6}$ to $x\approx 1$. 
The exponents $\alpha(x,z)$ found by such an analysis are displayed in Fig.~\ref{alphazxvslogx}. 
For very low values of $x<10^{-5}$, 
one observes that $w(x,z) \sim x^{1/2}$ for all the investigated range of $z$.
For $10^{-5}<x<10^{-1}$, the curves corresponding to $z$-positions close to the walls reach a $1/3$ power law regime. 
In this regime, the local exponent in the middle of the microchannel deviates from $1/2$, reaching values up to $0.55$.
Eventually, the classical $1/2$ exponent of homogenous diffusion is recovered for $x\gtrsim 0.1$.
  
At small but finite distance from the walls ($\lesssim 0.2$), 
we therefore predict a succession of three power law regimes with 
exponents $1/2$, $1/3$, and $1/2$.
The transition from the $1/2$ to the $1/3$ regime occurs earlier
for values of $z$ closer to the walls. To our knowledge, no experiments have shown this transition.

\begin{figure}[ht!]
\begin{center}
\scalebox{1}{\includegraphics{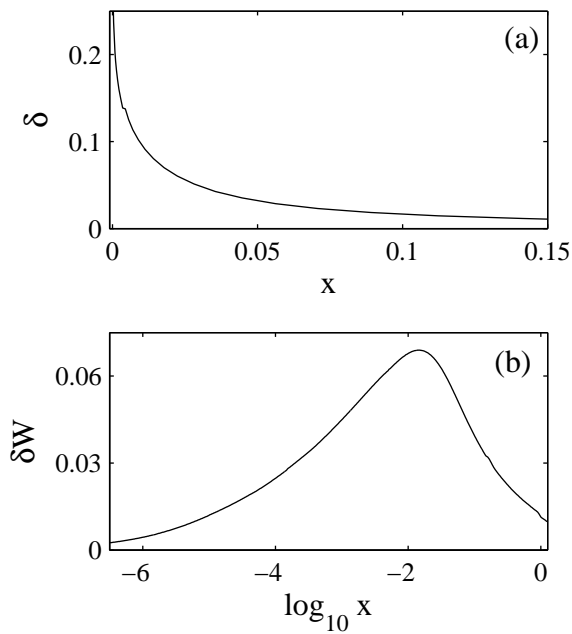}}
\end{center}
\caption{{\small (a) $\delta = w_m(x)/\sqrt{3x/2} - 1$ vs. $x$. $\delta$ allows one to estimate
the deviation between the width of the averaged concentration profiles and the 
solution  when the Poiseuille flow has not been taken into account.
(b) $\delta W$ vs. $x$. $\delta W$ 
is the difference between the widths of the diffusion zone at $z\rightarrow \pm 1/2$ and at $z=0$.
}} \label{largeur_moy.eps}
\end{figure}

In many experiments, the quantity measured is the concentration field averaged over the thickness of the microchannel. 
We compute this quantity from our solutions of Eq.~(\ref{eq:principale}),
and define the corresponding width $w_m(x)$ describing broadening 
in this averaged map using the analog of Eq.~(\ref{definition}).
We can now quantitatively answer one of the main points raised in the introduction:
for very thin channel, how good/bad is the approximation consisting
in assuming fast diffusion over $Z$ so that flow heterogeneities
can altogether be neglected ? This approximation leads 
to the homogeneous diffusion prediction $w(x) = \sqrt{3x/2}$.
We compare it to the real width of the averaged map  
$w_m(x)$ by plotting in Fig.~\ref{largeur_moy.eps}(a) 
a measure of the relative error made in using this approximation $\delta(x) = w_m(x)/\sqrt{3x/2} - 1$. 
From the plot it is clear that while important deviations are observed
at very short distances, the error is less than 5\% for all values of $x$
larger than $0.03$, which corresponds in real units to distances
after the junction $X \le 0.03\,\Pe\,d$.

Figure~\ref{largeur_moy.eps}(b) shows another measure of the
heterogeneity along $z$ of the concentration profile: $\delta W$,
the difference between the widths of the interdiffusion zone at 
the wall and at $z=0$. As can be seen on this plot, the maximal difference is reached for 
$x\approx 10^{-2}$, and corresponds to values of about 7\% (in real units $0.07\,d$). 
\subsection{Transition of the exponent from 1/2 to 1/3 close to the walls}

We present now analytical arguments 
that help understand the above results, namely the transition
from the $1/2$ to the $1/3$ regime observed close to the walls, 
for a value of $x$ that increases with the distance to the wall. 

We focus on the vicinity of the bottom wall, and make therefore 
the additional reasonable approximation that the velocity profile there is almost linear,  
so: $v(Z') = \gammap Z'$, 
where $\gammap = 4U/d$ is the local shear rate at the wall and $Z' = Z+d/2$ is the distance to the wall. In the 
diffusion process, the only relevant length scale is therefore $l = \sqrt{D/\dot{\gamma}}$.
Close to the wall, the concentration field of the solute A evolves through
\begin{equation}
Z'\partial_X a =  l^2 (\partial_Y^2 + \partial_{Z'}^2) a\,. \label{closetothewall}
\end{equation}
We make a change of variables akin to the classical treatment
of the L\'evêque problem: 
\begin{align}
u &= Z'/(X^{1/3}l^{2/3})\,, \\
v &= Y/(X^{1/3}l^{2/3})\,, \\
a &= F(u,v)\,, 
\end{align}
so that Eq.~(\ref{closetothewall}) reads
\begin{align}
(\partial_u^2 + \partial_v^2)F = -\frac{1}{3}u[u\partial_u F + v\partial_v F]\label{eq:armand} \,.
\end{align}
The boundary and initial conditions  
\begin{align}
&(\partial_{Z'} a)_{Z'=0} = 0\,, \\
&a = 0~~\text{for}~~Y>0,~~\text{and}~~a = 1~~\text{for}~~Y<0\,,
\end{align}
read now
\begin{align}
&(\partial_u F)_{u=0} = 0\,,\\
&\lim_{v\rightarrow + \infty} F = 1 \,,\\
&\lim_{v\rightarrow -\infty} F = 0 \,. 
\end{align}
Using Eq.~(\ref{definition}), the local width of the concentration profile is:
\begin{align}
w^2(X,Z') = X^{2/3} l^{4/3} G(u) \,, \label{maitresse}
\end{align} 
where
\begin{align}
&G(u) = \int \dd v\,v^2 \partial_v F \,.
\end{align} 
Using Eq.~(\ref{eq:armand}), $G(u)$ has to satisfy
\begin{align}
G'' + \frac{u^2}{3}G' - \frac{2 u}{3} G + 2 = 0\,, \label{eq:A}
\end{align} 
with
\begin{align}
&(\partial_u G)_{u=0}= 0\,,\\
&\lim_{u\rightarrow \infty} G = 0\,.
\end{align}

Close to the wall $Z' \sim 0$ and $u\sim 0$, the conditions $(\partial_u G)_{u=0}= 0$ therefore 
gives $G(u) = \cste$ and $w(X,Z) \sim X^{1/3}Z'^{2/3}$. This is the classical results of the Lévêque problem 
\cite{Stone:89} and
shown by our simulations for positions $Z'$ very close to the walls, for $x<0.1$.    
For large $u$, one should have $w^2(X,Z') = K D X/(\gammap Z')$, and therefore $G(u) = K/u$. Using Eq.~(\ref{eq:A}),
one has $K=2$ which is the classical diffusive behaviour. 

\begin{figure}[ht!]
\begin{center}
\scalebox{1}{\includegraphics{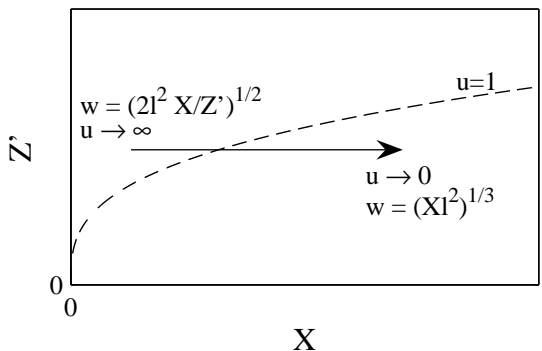}}
\end{center}
\caption{{\small Schematic representation of the asymptotic behaviour of the solutions of Eq.~(\ref{eq:A}).
The line $X=0$ corresponds to $u \sim \infty$ for finite $Z'$. The dashed line indicates $u=1$. \label{figure_armand}}}
\end{figure}

Figure~\ref{figure_armand} helps to sum up these results.
At a given position $Z'$, one encounters two regimes as $X$ is increased. For $X\sim 0$, one has 
$u\rightarrow \infty$ and $w(X,Z') \sim X^{1/2}$, and for large $X$ one has $u\rightarrow 0$ and $w(X,Z') \sim X^{1/3}$.
The boundary between these two regimes can be estimated using the curve $u=1$, i.e. 
$Z'=X^{1/3}l^{2/3}$. 
In the previous systems of units [see Eq.~(\ref{eq:adi})], $u=1$ reads $z = \pm 0.5 \mp (x/4)^{1/3}$. 
One of these two boundaries is shown in Fig.~\ref{alphazxvslogx} and a good agreement
is observed with the frontier between the $1/2$ and $1/3$ regimes.   

Note that this simple argument, that permits to apprehend the transition from 
1/2 to 1/3, is strictly speaking limited to the vicinity of the walls.
Further  away, one should take into account the curvature 
of the flow profile, and eventually describe the merger of the 
two zones of $1/3$ behaviour.

\subsection{Validity of the approximations embedded in the model}

We have altogether neglected longitudinal diffusion
(along $X$) on the ground of a large P\'eclet number
 $\Pe=Ud/D \gg1$. To assess more locally the range of validity of this approximation,
 a useful quantity is $v(Z)X/D$, the P\'eclet number using the length scale $X$ and the velocity $v(Z)$. 
Indeed, positions $X,Z$ such that $v(Z) X /D \gg 1$ correspond to the region where diffusion along 
the axial direction is negligible. The curve $v(Z) X / D = 1$ is thus a 
reasonable estimate for the boundary of the domain where our approximation should be valid.
In non-dimensionalized units, this boundary reads
\begin{eqnarray}
x = \frac{1}{\Pe^2}\frac{1}{1-(2z)^2} \label{eq:validity}\,,
\end{eqnarray}     
and is plotted in Fig.~\ref{alphazxvslogx} for the case $\Pe = 1000$. 
It intersects the curve $u=1$ indicating the growing boundary layer 
corresponding to the $1/3$ power law regime for values $x \approx 1/\Pe^2$ 
and $z \pm 1/2 \approx \pm (1/4\Pe^2)^{1/3}$. So, to visualize experimentally 
comfortably this 1/3 regime, one should use high P\'eclet numbers, and 
focus on positions $(z, x)$ given by the previous relations. 
For $\Pe=1000$, the crossing point (see Fig.~\ref{alphazxvslogx}) 
is at $z \pm 1/2 \approx 6\,10^{-3}$, i.e. below 1\% of the height of the microchannel, 
leaving ample space for observing this regime.   
 
Another issue concerns having neglected any specific effect in the entrance
region of size $L$. In this zone, the velocity profile evolves from 
what it was in each branch into its final form given Eq.~(1).
 The extent of this region in non-dimensional units is of the order of
 $x \lesssim (L/d)\,\Pe^{-1}$, so that unless $\Pe$ is really very large,
 this region will overlap with the $1/3$-regime. For example for $\Pe=1000$
 and $L/d=10$, the entrance region corresponds to $x \lesssim 0.01$.
 However, we do not think that a refined description with a model for this entrance 
 region would modify the picture presented. Indeed, it takes only a distance of order 
 $X\sim d$ after the apex of the junction to reach a parabolic profile in the interfacial
 region, and it is only the amplitude of this parabola (the local maximal velocity) 
that relaxes to its asymptotic value over $X\sim L$.
 
\subsection{Comparison with experiments}
We now compare our results with existing experimental data.
Kamholz {\it et al.} have measured the diffusivity of various fluorescent molecules
using the device sketched in Fig.~\ref{setup} \cite{Kamholz:01_1}. 
They used microchannels of small thickness ($d=10~\mu$m)
and models assuming fast averaging over this thin dimension
to extract the diffusion constants.
Our analysis shows that this procedure was actually
appropriate given its simplicity. Indeed, their "worst" experimental configuration
(highest flow rate $\approx 1000~$nL/s, and smallest diffusivity $D=6.2\,10^{-11}~$m$^2$/s 
for streptavidin, so $\Pe \approx 8000$), 
corresponds to measurements taken at $x\approx 0.05$ in non-dimensionalized units,
for which the relative error from using the simplified model 
is less than 5\% as can be seen on Fig.~\ref{largeur_moy.eps}(a)
(even if the difference of width $\delta W$ is maximal at this position [see Fig.~\ref{largeur_moy.eps}(b)]).

As pointed in the introduction of the present paper,
our results are not consistent with the theoretical data discussed in Ref.~\cite{Kamholz:02}.
In this work, the authors do no use non-dimensionalized equations but have
computed the concentration profiles for a  
specific case (flow rate $Q=42~$nL/s, $L=2405~\mu$m, $d=10~\mu$m, $D=3.4~10^{-10}~$m$^2$\,s$^{-1}$, 
leading to $\Pe \approx 50$). They found, 
in the midplane of the channel ($z=0$)  and in the range $x=10^{-2}$--$10^{-1}$, 
a power law regime for the transverse spreading 
given by an exponent 2/3. Such a result is not consistent with the data displayed in Fig.~\ref{alphazxvslogx}, and we 
believe it is due to the fitting procedure of the curves $w(x,z)$ vs. $x$.
More important, the authors claim that a model with a homogeneous velocity  profile (1D model) overpredicts 
the diffusion width by $\approx 35\%$, even for distances $X=1000~\mu$m, i.e. $x=2$ in non-dimensionalized units.
Such results are in contradiction with our data [see Fig.~\ref{largeur_moy.eps}(a)].

We now consider the data of Ismagilov {\it et al.}~\cite{Ismagilov:00},
who used confocal fluorescence microscopy to measure the transverse diffusive broadening 
of a fluorescent probe formed as the product  of a reaction
involving diffusing reactants originating from solutions flowing side by side in a microchannel. 
Clearly, that situation is more complex than the one analyzed above
as both diffusion and reaction come into play.  
The data displayed in Ref.~\cite{Ismagilov:00}, more precisely in their Fig.~2,  concern 
a microchannel of thickness $d=105~\mu$m, a maximal velocity of the order of $12$~cm/s, and an investigated
range of $X$ ranging between $10^2$ and $10^4~\mu$m. 
Taking $D\approx 10^{-9}$~m$^2$/s to estimate $\Pe \approx 12600$, 
the dimensionless $x$ ranges roughly between $10^{-4}$ and $10^{-2}$. 
As can be seen on Figs.~\ref{logwvslogx} and~\ref{alphazxvslogx}, 
our analysis suggests that in this range, 
the boundary layers where the transverse diffusive zone widens as  $x^{1/3}$
should occupy a significant portion of the microchannel,
in agreement with the observation by the authors of such a scaling
for the broadening close to the walls.   
However, as can be seen on Fig.~\ref{largeur_moy.eps}(b), 
from our analysis, the difference between diffusive width at the walls
and at the center should be less than $\approx 0.07 d$, i.e. $\approx 7~\mu$m. 
The values reported in Ref.~\cite{Ismagilov:00} lead to a somewhat
larger estimate ($\approx 30~\mu$m)-- the difference is too large
to be simply the result of a slightly different definition of the diffusion width $w$  
[see Eq.~(\ref{definition})]--. 
Furthermore, the interdiffusion zones they observed have a different shape than those displayed in Fig.~\ref{cartes}. 
We conclude that while the presence and location of a "1/3 regime"
is robust to the differences between the two situations,
some features do show a difference. To confirm this we now proceed
to a brief investigation of a reaction-diffusion model.
        
\section{Diffusion of quickly reacting species}

\subsection{Assumptions of the model}

We turn to the case of a chemical reaction $\text A+\text B \rightleftarrows {\text{C}}$. 
Different groups have theoretically investigated the solutions of the unidimensional problem of
reaction-diffusion without advection \cite{Galfi:88,Taitelbaum:92,Bazant:00}. These works mainly deal
with irreversible reactions  $\text A+\text B \rightarrow {\text{C}}$, and have provided scaling laws 
of the width of the reaction front, both in the asymptotic and in the short-time limits.
In our case, we are interested in the concentration map of the product C, when the two reactants, A and~B, are 
injected in the two arms of the microchannel (see Fig.~\ref{setup}), at a same given constant flow rate, and at
concentrations $a_0$ and $b_0$ respectively.
As above, we restrict ourselves to dilute
solutions, and further focus on fast chemical reactions, so that
 chemical equilibrium $AB/C = K_{eq}$ is satisfied at each point of the microchannel
($K_{eq}$ is the equilibrium constant of the reaction, and $A$, $B$ and $C$ 
are the concentrations of species A, B, and  C respectively).
To reduce the number of independent parameters, we take the diffusion coefficients of 
species {B} and {C} to be  almost identical, i.e. $D_C\approx D_B$. 
This should well describe complexation reactions where A is a small molecule 
that can not affect the diffusion coefficient of the larger  B. 
Reaction-diffusion dynamics in a microchannel with the same 
assumptions than previously read
\begin{eqnarray}
v(Z)\partial_X A &= -R + D_A(\partial_Y^2 + \partial_Z^2)A \label{eq:11}\,, \\ 
v(Z)\partial_X B &= -R + D_B(\partial_Y^2 + \partial_Z^2)B\,, \notag \\
v(Z)\partial_X C &= +R + D_B(\partial_Y^2 + \partial_Z^2)C\,,  \notag
\end{eqnarray}
$R$ corresponds to the reaction term.
Since the reaction is very fast, $AB/ C = K_{eq}$, and the problem is 
governed by only two equations. Considering species conservation, we focus on 
the combination: $P=A+C$, and $Q=B+C$, so Eqs.~(\ref{eq:11}) read
\begin{eqnarray}
v(Z)\partial_X P &=& D_A(\partial_Y^2 + \partial_Z^2)P \label{eq:PQ}\\ 
&+& (D_A - D_B) (\partial_Y^2 + \partial_Z^2)C \,, \notag \\ 
v(Z)\partial_X Q &=& D_B(\partial_Y^2 + \partial_Z^2)Q\,, \notag 
\end{eqnarray} 
where $C=C(P,Q)$ is to be understood as the implicit solution
the local chemical equilibrium:
 \begin{eqnarray}
K_{eq} C = (P-C)(Q-C) \label{eq:poly}\,. 
\end{eqnarray} 

We move to dimensionless variables,
\begin{eqnarray}
p = P/a_0\,,~~~~~q = Q/b_0\,,~~~~c = C/(a_0 b_0)^{1/2}\,,
\end{eqnarray}
 and again~:
\begin{eqnarray}
x = X/(d\,\Pe), \,\,\,\,\, z = Z/d, \,\,\,\, \text{and} \,\,\,\, y = Y/d\,, 
\end{eqnarray}
with $\Pe = U d /D_A$, 
Eqs.~(\ref{eq:PQ}) and (\ref{eq:poly}) read
\begin{eqnarray}
&&(1-(2z)^2)\,\partial_x p = (\partial_y^2 + \partial_z^2) (p + \frac{\chi^2-1}{\beta}c)\,,\label{eq:reac_adi}\\
&&(1-(2z)^2)\,\partial_x q = \chi^2 (\partial_y^2 + \partial_z^2) q\,, \notag \\
&&c^2 - c(\gamma + \beta p + q/\beta) + pq = 0 \notag \,,
\end{eqnarray} 
where we now have {\it three} dimensionless parameters that control the outcome:
\begin{eqnarray}
\chi^2 = D_B/D_A,~~\beta^2 = a_0/b_0,~~\gamma = K_{eq}/(a_0 b_0)^{1/2}\,.
\end{eqnarray}
As can be seen on Eqs.~(\ref{eq:reac_adi}), the case $\chi=1$ 
yields $p=q$ and is equivalent to the problem discussed previously. 
In general $\chi \neq 1$ and $\beta \neq 1$, so the diffusion pattern is not symmetric.

\subsection{Numerical computation and results}
As above, we write Eqs.~(\ref{eq:reac_adi}) on a discrete set of $n=100\times p = 80$~points.
 We impose the classical no-flux 
boundary conditions at all the walls of the microchannel, and the 
initial conditions at $X=0$ are $p=1~~(q=0~~\text{resp.})
~~\text{for}~~Y>0$, and $p=0~~(q=1~~\text{resp.})~~\text{for}~~Y<0$.  

\begin{figure}[ht!]
\begin{center}
\scalebox{1}{\includegraphics{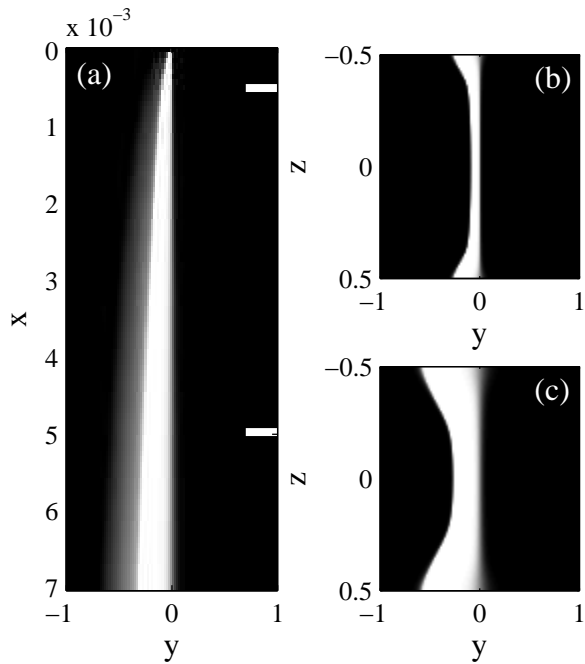}}
\end{center}
\caption{{\small  
(a) concentration field $c$ of the product C averaged over the height of the microchannel in the case of a 
chemical reaction described by Eqs.~(\ref{eq:reac_adi})  with 
 $\chi = 0.3$, $\beta=\sqrt{200}$, and $\gamma=0.0055$. (b-c) two slices $y$ vs. $z$ of the concentration of the 
product of the reaction, at the locations specified by white lines on (a).  
\label{carte3D.eps}}}
\end{figure}

Figure~\ref{carte3D.eps} shows the
results obtained for a specific set of parameters $\chi = 0.3$, $\beta=\sqrt{200}$, 
and $\gamma=0.0055$, chosen because it 
corresponds roughly to the situation studied by Ismagilov {\it et al}. 
in Ref.~\cite{Ismagilov:00} ($D_A \approx 1.2\,10^{-9}$, 
$D_B \approx 10^{-10}$~m$^2$\,s$^{-1}$, $a_0=1~$mM, $b_0=5~\mu$M, $K_{eq} = 0.39~\mu$M). 
The depth-averaged concentration map $c$ in the microchannel is displayed
in this figure (flow is downwards), together with two cross-sectional plots
of the concentration in the ($y$,$z$) plane.
As for the diffusion of a solute, the diffusion zone is not homogeneous over the height of the channel, due to the 
presence of the Poiseuille profile. Moreover, the diffusion zone is not symmetric 
($x$,$y$), as a consequence of the stoechiometric imbalance ($\beta \neq 1$), 
and the asymmetry in diffusion coefficients ($\chi \neq 1$) \cite{Salmon:05_1}.
The good agreement between the 
experimental shapes of the concentration profiles 
obtained in Ref.~\cite{Ismagilov:00} and our simulation is clear.
Moreover, the differences $\delta W$ between the widths of the interdiffusion zone at 
the wall and at $z=0$, are now consistent with the experimental data 
(for example, in Fig.~2 of Ref.~\cite{Ismagilov:00}, $\delta W \approx 0.1$--$0.3$, for $\Pe \approx 12600$ and 
$x\approx$ 0.0001--0.001, our simulations give roughly the same results).

\begin{figure}[ht!]
\begin{center}
\scalebox{1}{\includegraphics{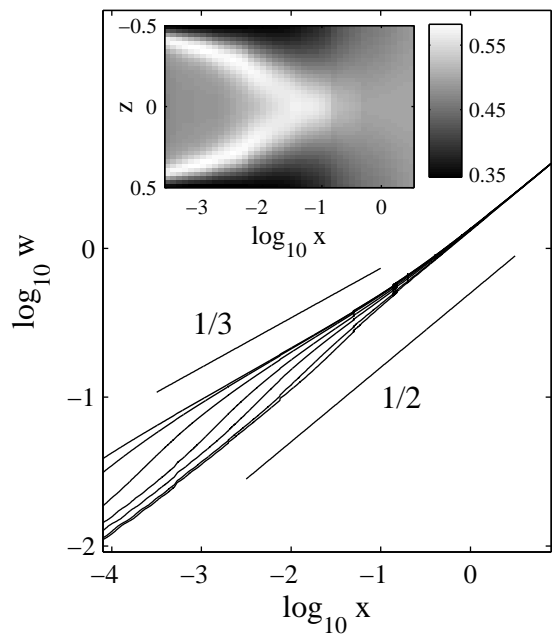}}
\end{center}
\caption{{\small
 $\log_{10}(w)$ vs. $\log_{10}(x)$ for different $z$ in the case of Eqs.~(\ref{eq:reac_adi}) with 
 $\chi = 0.3$, $\beta=\sqrt{200}$, and $\gamma=0.0055$.
The two continous lines indicate the $1/2$ and $1/3$ power law behaviours. Insert: corresponding 
exponents $\alpha(x,z)$ vs. $\log_{10}(x)$ and $z$, obtained by local fits of $w$ vs. $x$ by power laws. 
\label{analyse_reac_chimique}}}
\end{figure}

As above (see Figs.~\ref{logwvslogx} and \ref{alphazxvslogx}), 
we perform numerical simulations of Eqs.~(\ref{eq:reac_adi}) over a wide range of $x$, and compute $w(x,z)$. 
The corresponding results are shown in Fig.~\ref{analyse_reac_chimique}, 
together with a map  of the local exponents $\alpha(x,z)$,
estimated by fitting  the curves over a sliding decade of $x$ with a law $w(x,z) \sim x^{\alpha(x,z)}$ 
(in this case we have chosen $n=80$ and $p=40$ for the simulations).
Obviously the phenomenology is very similar to that for the simple
diffusion of a solute studied earlier: (i)
close to the walls at $z = \pm 0.5$, $w\sim x^{1/3}$, whereas for $z\approx 0$,  then $w\sim x^{1/2}$; (ii)
beyond $x\approx 0.1$, the diffusion zone is homogeneous over the height of the channel and 
widens as $w\sim x^{1/2}$; (iii)  close to the walls
the sequence of power law regimes with exponents $1/2$, $1/3$, and $1/2$ is recovered; (iv) the locus of the transition 
from the $1/2$ to the 1/3 regime occurs earlier for $z$ closer to the walls.
We find similar results when solving numerically
Eqs.~(\ref{eq:reac_adi}) for different set of parameters $\chi$, $\beta$ and $\gamma$.
However, the increased complexity of the equations does not permit to perform 
as easily as in subsection II-D a simple analytical argument to formally
argue for this phenomenology to hold whatever the values of the parameters.  
Theoretical approaches such as the ones described in Refs.~\cite{Galfi:88,Taitelbaum:92,Bazant:00}
may be useful to answer these questions. 

\section{Conclusion}

The present numerical and theoretical study shows the generality
of features reported for specific conditions by previous authors
on lateral transport of solutes between streams in pressure-driven flows.
For diffusion alone, the variations of the velocity along the thickness of the channel 
lead to remarkable features: while in the midplane 
of the microchannel the diffusive width classically grows with a $1/2$ exponent,
close to the wall the broadening downstream can go through three power law
regimes with exponents $1/2$, $1/3$ and $1/2$ again. Remarkably, our numerical computations
suggest that this phenomenology may also hold for reaction-diffusion problems,
at least in the diffusion-controlled limit of fast reaction kinetics.
All our discussion and arguments, together with the various plots presented,
are proposed in non-dimensional form, which hopefully should help
experimentalists in assessing a priori the relevance of 3D effects
in microfluidic experiments involving the lateral transport of solutes.

\acknowledgments{J.-B. S. thanks P. Tabeling and all the members of the microfluidic group at 
ESPCI (MMN, UMR 7083) for many fruitful discussions.}

\end{document}